\documentclass[conference]{IEEEtran}
\IEEEoverridecommandlockouts
\usepackage{cite}
\usepackage{amsmath,amssymb,amsfonts}
\usepackage{algorithmic}
\usepackage{graphicx}
\usepackage{textcomp}
\usepackage{xcolor}
\usepackage{pgfplots}
\usepackage{pgfplotstable}
\usepackage{tikz}
\def\BibTeX{{\rm B\kern-.05em{\sc i\kern-.025em b}\kern-.08em
    T\kern-.1667em\lower.7ex\hbox{E}\kern-.125emX}}
\begin{document}

\title{Energy-Efficient Software–Hardware Co-Design for Machine Learning: From TinyML to Large Language Models
}

\author{\IEEEauthorblockN{1\textsuperscript{st} Mohammad Saleh Vahdatpour}
\IEEEauthorblockA{\textit{Department of Computer Science} \\
\textit{Georgia State University}\\
Atlanta, GA, USA \\
mvahdatpour1@student.gsu.edu}
\and
\IEEEauthorblockN{2\textsuperscript{nd} Yanqing Zhang}
\IEEEauthorblockA{\textit{Department of Computer Science} \\
\textit{Georgia State University}\\
Atlanta, GA, USA \\
yzhang@gsu.edu}
}

\maketitle

\begin{abstract}
The rapid deployment of machine learning across platforms—from milliwatt-class TinyML devices to large language models—has made energy efficiency a primary constraint for sustainable AI. Across these scales, performance and energy are increasingly limited by data movement and memory-system behavior rather than by arithmetic throughput alone. This work reviews energy-efficient software--hardware co-design methods spanning edge inference and training to datacenter-scale LLM serving, covering accelerator architectures (e.g., ASIC/FPGA dataflows, processing-/compute-in-memory designs) and system-level techniques (e.g., partitioning, quantization, scheduling, and runtime adaptation). We distill common design levers and trade-offs, and highlight recurring gaps including limited cross-platform generalization, large and costly co-design search spaces, and inconsistent benchmarking across workloads and deployment settings. Finally, we outline a hierarchical decomposition perspective that maps optimization strategies to computational roles and supports incremental adaptation, offering practical guidance for building energy- and carbon-aware ML systems.
\end{abstract}


\section{INTRODUCTION}

The rapid expansion of artificial intelligence across natural language processing, computer vision, and autonomous systems has introduced significant sustainability challenges. Training and deploying modern machine learning models incur substantial environmental and operational costs, ranging from battery-powered edge inference to megawatt-scale datacenter workloads. Training a single large language model can emit carbon comparable to multiple passenger vehicles over their lifetimes \cite{fernandez2025energy}, while billions of edge devices must operate under stringent energy constraints \cite{solanki2025atm, husom2025sustainable}. Recent studies further indicate that inference now accounts for more than half of total LLM lifecycle emissions \cite{jegham2025hungry}, underscoring the urgency of energy-efficient deployment strategies across the full spectrum of machine learning systems.

A central contributor to this challenge is the memory wall, where data movement between compute units and memory dominates both latency and energy consumption. DRAM accesses can consume up to 200$\times$ more energy than arithmetic operations \cite{Chen2016Eyeriss, kim2025hardware}, an imbalance that intensifies as models scale from kilobyte-level networks on microcontrollers to trillion-parameter foundation models \cite{wolters2024memory, yildirim2025risc}. For TinyML systems, even milliwatt-level inefficiencies can render deployments impractical \cite{solanki2025atm}, while large-scale LLM infrastructures require thousands of accelerators, leading to megawatt-scale power draw and substantial operational costs \cite{fernandez2025energy, maliakel2025investigating}.

Traditional approaches that optimize machine learning algorithms independently of their execution platforms are increasingly inadequate \cite{Bringmann2021EdgeAI, Guo2025LLMSurvey}. Software--hardware co-design has therefore emerged as a key paradigm, jointly optimizing models, compilers, architectures, and runtime systems to improve energy efficiency, throughput, and adaptability \cite{Rashidi2023UNICO, Moreau2019Blueprint}. By coordinating decisions across the computational stack, co-design enables cross-layer optimizations that are difficult or impossible to achieve through isolated improvements \cite{Bringmann2021EdgeAI}.

The resulting co-design landscape is highly heterogeneous, as illustrated in Fig.~\ref{fig:timeline}. TinyML accelerators leverage FPGA-based architectures and zero-buffer dataflows to minimize intermediate memory accesses \cite{solanki2025atm, yildirim2025risc, Chu2023GGNN}, while energy-harvesting systems further improve efficiency through adaptive precision and early termination mechanisms \cite{solanki2025atm}. Edge--cloud deployments employ split inference strategies to balance latency and energy across distributed resources \cite{may2024dynasplit}. At larger scales, transformer and LLM accelerators increasingly adopt compute-in-memory techniques to mitigate data movement overhead \cite{kim2025hardware, Zhou2022TransPIM, yu2025digital}, and FPGA-based spatial accelerators demonstrate competitive inference performance within constrained power envelopes \cite{li2025design, chen2024understanding}.

Despite this progress, several challenges remain unresolved. Many co-design techniques lack generalization across model families and deployment scales, with approaches optimized for convolutional networks often failing to transfer effectively to attention-based architectures \cite{Guo2025LLMSurvey, kachris2025survey}. Moreover, the co-design search space—spanning dataflows, quantization schemes, memory hierarchies, and parallelism strategies—requires extensive manual tuning or specialized expertise \cite{Bringmann2021EdgeAI, Rashidi2023UNICO}. Existing efforts rarely provide unified frameworks that bridge optimization strategies across TinyML, edge, and LLM systems \cite{kachris2025survey}, while benchmarking practices frequently rely on idealized metrics that underestimate real-world energy consumption and operational constraints \cite{fernandez2025energy, maliakel2025investigating}.

Prior surveys have examined transformer accelerators \cite{kachris2025survey}, LLM-oriented co-design \cite{Guo2025LLMSurvey}, and FPGA-based neural network implementations, yet comprehensive analysis spanning the full machine learning spectrum remains limited. Existing reviews often focus on either datacenter-scale systems or edge-only deployments, missing insights enabled by cross-scale comparison, and sustainability considerations are frequently secondary to throughput-centric evaluation \cite{fernandez2025energy, jegham2025hungry}.

In this work, we synthesize energy-efficient software--hardware co-design methodologies spanning sub-milliwatt TinyML systems \cite{solanki2025atm, yildirim2025risc, Behnam2023SUSHI} to datacenter-scale LLM infrastructures \cite{fernandez2025energy, Lie2023Cerebras}. We draw insights from FPGA accelerators \cite{li2025design, chen2024understanding, Zhang2018DNNBuilder}, compute-in-memory architectures \cite{kim2025hardware, wolters2024memory, Zhou2022TransPIM, yu2025digital, ma2025hnm}, ASIC-based designs \cite{Rashidi2023UNICO, Chen2019EyerissV2}, and hybrid platforms \cite{Ambrosi2018AnalogDigital, Fan2022M3ViT}. Through comparative analysis, we identify recurring design trade-offs and gaps that hinder practical deployment of sustainable AI systems, and discuss hierarchical decomposition with FPGA-based incremental learning as a unifying perspective for efficient and adaptable training and inference across deployment scales \cite{vahdatpour2025energy}. The remainder of the paper establishes key background concepts, surveys co-design techniques across system scales, and synthesizes open challenges and future research directions toward carbon-aware AI infrastructure.

\begin{figure}[t]
\centering
\begin{tikzpicture}
\begin{axis}[
    ytick={2016,2017,...,2025},
    yticklabels={'16,'17,'18,'19,'20,'21,'22,'23,'24,'25},
    ylabel={\small{Year}},
    ylabel style={yshift=-10pt},
    y dir=normal,
    xtick=data,
    xticklabel style={rotate=45, anchor=east, font=\tiny},
    xticklabels={
        {Eyeriss},
        {DNNBuilder},
        {Cambricon-S},
        {Analog-Dig.},
        {Eyeriss v2},
        {TVM Unity},
        {EdgeML},
        {HASCO},
        {SECDA},
        {TransPIM},
        {M³ViT},
        {GGNN},
        {Cerebras},
        {UNICO},
        {SUSHI},
        {CiM Secure},
        {LLM-CiM},
        {Fire-Flyer},
        {DynaSplit},
        {LLM Quant},
        {HASTILY},
        {DSC Accel},
        {ATM-Net},
        {HNM-CiM},
        {Attn-CiM},
        {FPGA-Trans},
        {LLM DVFS},
        {vLLM Opt},
        {Hierarchical}
    },
    height=6cm,
    width=\columnwidth,
    enlargelimits=0.05,
    grid=major,
    grid style={dashed, gray!20},
    legend style={
        at={(0.59,0.52)}, 
        anchor=north west, 
        legend columns=2, 
        font=\tiny,
        /tikz/every even column/.append style={column sep=0.3cm}
    },
    scatter/classes={
        ASIC={mark=*,blue!80!black},
        FPGA={mark=square*,green!70!black},
        TinyML={mark=triangle*,orange!90!black},
        PIM={mark=diamond*,purple!80!black},
        CiM={mark=pentagon*,red!80!black},
        Compiler={mark=*,cyan!70!black},
        Wafer={mark=square*,magenta!80!black},
        Analog={mark=otimes*,gray!60!black},
        EdgeCloud={mark=star,brown!70!black},
        GPU={mark=Mercedes star,teal!70!black},
        Datacenter={mark=10-pointed star,violet!70!black}
    }
]
\addplot[scatter, only marks, scatter src=explicit symbolic, mark size=2.5pt]
    coordinates {
        (0,2016) [ASIC]         
        (1,2018) [FPGA]         
        (2,2018) [ASIC]         
        (3,2018) [Analog]       
        (4,2019) [ASIC]         
        (5,2019) [Compiler]     
        (6,2020) [TinyML]       
        (7,2021) [ASIC]         
        (8,2021) [FPGA]         
        (9,2022) [PIM]          
        (10,2022) [FPGA]        
        (11,2023) [FPGA]        
        (12,2023) [Wafer]       
        (13,2023) [ASIC]        
        (14,2023) [TinyML]      
        (15,2023) [CiM]         
        (16,2023) [CiM]         
        (17,2024) [Datacenter]  
        (18,2024) [EdgeCloud]   
        (19,2024) [GPU]         
        (20,2025) [CiM]         
        (21,2025) [FPGA]        
        (22,2025) [FPGA]        
        (23,2025) [CiM]         
        (24,2025) [CiM]         
        (25,2025) [FPGA]        
        (26,2025) [GPU]         
        (27,2025) [GPU]         
        (28,2025) [FPGA]        
    };
\legend{ASIC, FPGA, TinyML, PIM, CiM, Compiler, Wafer-Scale, Analog, Edge-Cloud, GPU Opt, Datacenter}
\end{axis}
\end{tikzpicture}
\caption{Timeline of representative software–hardware co-design approaches for energy-efficient machine learning (2016–2025), categorized by architectural paradigm.}
\label{fig:timeline}
\end{figure}

\section{BACKGROUND}
\label{sec:background}

\subsection{The Memory Wall and Data Movement Bottleneck}

A fundamental challenge in modern machine learning accelerators arises from the von Neumann separation of computation and memory. While computational throughput has scaled rapidly, memory bandwidth and access efficiency have lagged behind, giving rise to the well-known memory wall \cite{wolters2024memory, kim2025hardware}. Empirical studies consistently show a steep energy hierarchy: off-chip DRAM accesses consume roughly 200$\times$ more energy than multiply-accumulate operations, while on-chip SRAM access requires approximately 10$\times$ the computation energy \cite{Chen2016Eyeriss}. As model sizes grow from kilobyte-scale neural networks to billion- and trillion-parameter models, repeated movement of weights and activations increasingly dominates both latency and energy consumption \cite{yildirim2025risc, kim2025hardware}.

Early accelerator designs such as Eyeriss demonstrated that careful dataflow and memory hierarchy optimization can significantly mitigate this bottleneck by maximizing data reuse and minimizing off-chip access \cite{Chen2016Eyeriss, Chen2019EyerissV2}. However, attention-based architectures intensify memory pressure. Self-attention exhibits quadratic complexity in sequence length, producing large intermediate matrices that must be stored and accessed multiple times \cite{Zhou2022TransPIM, kim2025hardware, yu2025digital}. For long-context LLM inference, these activations often exceed on-chip capacity, forcing frequent off-chip transfers and exacerbating the memory wall \cite{wolters2024memory, Zhou2022TransPIM}.

To address these challenges, recent work explores dataflow-level and architectural strategies that reduce or bypass intermediate memory traffic. Zero-buffer and fused dataflow designs compute outputs through tightly coupled pipelines without materializing intermediate feature maps, substantially reducing data movement \cite{yildirim2025risc}. Compute-in-memory approaches further challenge the von Neumann paradigm by performing arithmetic operations directly within memory arrays, colocating storage and computation to reduce access energy \cite{kim2025hardware, yu2025digital, ma2025hnm, wolters2024memory}.

\subsection{Scale-Dependent Efficiency Challenges}

Energy efficiency constraints vary substantially across deployment scales. TinyML systems target microcontrollers with limited on-chip memory—often below 256~KB SRAM—and operate under milliwatt-level power budgets \cite{Behnam2023SUSHI, yildirim2025risc}. In battery-powered and energy-harvesting IoT devices, inefficiencies directly translate into reduced lifetime or impractical maintenance costs, motivating adaptive architectures that dynamically adjust precision and computation depth based on available power \cite{solanki2025atm}. FPGA-based implementations are particularly effective in this regime, enabling reconfigurable dataflows tailored to specific model structures \cite{yildirim2025risc, Chu2023GGNN, Haris2021SECDA, Behnam2023SUSHI}.

Edge--cloud systems introduce a different set of trade-offs by partitioning computation between resource-constrained edge devices and centralized cloud backends \cite{may2024dynasplit}. Optimal partitioning must jointly consider local computation energy, communication overhead, and quality-of-service constraints. Prior work shows that carefully chosen split points can significantly reduce total energy consumption relative to cloud-only execution, although effectiveness depends strongly on model architecture and workload characteristics \cite{may2024dynasplit}. Prior studies further indicate that hardware acceleration at the edge can improve energy efficiency by reducing execution time, even when instantaneous power consumption is higher \cite{may2024dynasplit, vahdatpour2025forecasting}.

Large language models impose the most extreme energy demands during both training and inference \cite{fernandez2025energy, jegham2025hungry, maliakel2025investigating}. Models with billions of parameters exceed the capacity of single accelerators, requiring parallelism strategies that distribute computation and memory across multiple devices \cite{Lie2023Cerebras, Guo2025LLMSurvey}. Attention mechanisms further exacerbate memory pressure, particularly for long-context inference \cite{kim2025hardware}. Empirical benchmarking demonstrates that energy consumption varies significantly with task type, sequence length, and runtime behavior, and that system-level techniques such as DVFS, quantization, and inference serving optimizations can substantially reduce energy usage without modifying model semantics \cite{maliakel2025investigating, husom2025sustainable, fernandez2025energy}. At scale, these operational costs are compounded by infrastructure-level factors such as cooling and water usage, which contribute significantly to the overall environmental footprint of LLM deployments \cite{jegham2025hungry}.

\section{SOFTWARE-HARDWARE CO-DESIGN}
\label{sec:codesign}

\subsection{Co-Design Methodologies}

Software--hardware co-design departs from traditional isolated optimization by jointly developing machine learning algorithms and hardware architectures to optimize system-level objectives such as energy efficiency, performance, and accuracy \cite{Bringmann2021EdgeAI, Guo2025LLMSurvey}. Rather than treating hardware as a fixed execution substrate, co-design frameworks explore algorithmic and architectural parameters simultaneously, enabling cross-layer optimizations that are difficult to achieve through independent improvements \cite{Rashidi2023UNICO}. Representative approaches include Bayesian co-search techniques that jointly explore neural architectures and hardware configurations to improve robustness across energy-performance trade-offs \cite{Rashidi2023UNICO}, task-specialized tensor accelerators with configurable datapaths for fine-grained scheduling and memory reuse \cite{Xiao2021HASCO}, and sparse accelerator designs that exploit structured sparsity through specialized indexing and load-balancing mechanisms \cite{Zhou2018CambriconS}. Learned co-design further automates this process by incorporating latency, power, and area constraints directly into neural architecture search objectives \cite{Shi2020LearnedCoDesign}.

Despite these advances, automation remains a central challenge. The design space spans quantization precision, dataflow selection, tiling strategies, and parallelism configurations, making exhaustive manual exploration prohibitively expensive \cite{Bringmann2021EdgeAI, Danopoulos2024Survey}. Toolchains such as DNNBuilder provide automated mapping from high-level model descriptions to FPGA implementations, but are typically constrained to specific architecture classes \cite{Zhang2018DNNBuilder}. Compiler-based approaches, exemplified by TVM Unity, extend co-design through hardware-aware scheduling and adaptive tuning across heterogeneous platforms \cite{Moreau2019Blueprint}. However, such systems face limitations when targeting emerging workloads such as transformers, which exhibit irregular computation and memory access patterns \cite{kachris2025survey}.

\subsection{Cross-Layer Optimization Techniques}

Effective co-design requires coordinated optimization across algorithmic, architectural, and system layers. At the algorithm level, quantization reduces precision of weights and activations to lower-bit representations, substantially decreasing memory footprint and enabling efficient fixed-point execution, with quantization-aware training mitigating accuracy degradation \cite{solanki2025atm, husom2025sustainable}. Pruning techniques remove redundant parameters through structured or unstructured sparsity, while hybrid N:M sparsity patterns balance compression efficiency with hardware regularity \cite{ma2025hnm}. Knowledge distillation further compresses models by transferring representations from large teacher networks to compact student models \cite{Behnam2023SUSHI}.

At the architecture level, dataflow and memory hierarchy design play a dominant role in energy efficiency. Weight-stationary and output-stationary dataflows prioritize reuse of parameters or partial sums to reduce memory accesses \cite{Chen2016Eyeriss, yildirim2025risc}, while designs with minimal local buffering favor throughput at the cost of increased memory traffic \cite{yildirim2025risc}. Memory hierarchy optimization places frequently accessed data in energy-efficient storage tiers \cite{Chen2019EyerissV2}. Compute-in-memory architectures further reduce data movement by performing arithmetic directly within SRAM or emerging memory arrays, effectively collapsing the memory hierarchy \cite{kim2025hardware, yu2025digital, Zhou2022TransPIM}. Specialized attention accelerators address inefficiencies of conventional systolic arrays during softmax and normalization; examples include SRAM-based designs integrating exponential lookup within memory arrays \cite{kim2025hardware} and DRAM-based approaches performing attention score computation inside memory subarrays \cite{Zhou2022TransPIM}.

At the system level, runtime adaptation mechanisms align computation with dynamic operating conditions. Energy-aware scheduling adjusts precision or execution paths based on available power \cite{solanki2025atm}, while dynamic voltage and frequency scaling (DVFS) can reduce LLM inference energy by up to 30\% without altering model parameters \cite{maliakel2025investigating}. Split computing partitions neural networks across edge and cloud resources to jointly optimize local computation and communication costs \cite{may2024dynasplit}. Serving-level optimizations, including continuous batching, speculative decoding, and KV-cache management, improve throughput and efficiency for transformer inference, though their benefits depend on workload characteristics and latency constraints \cite{fernandez2025energy, may2024dynasplit}.

\section{ENERGY-EFFICIENT CO-DESIGN ACROSS THE ML SPECTRUM}
\label{sec:survey}

\subsection{TinyML: Ultra-Low-Power Edge Intelligence}

TinyML systems operate under extreme constraints, including sub-256KB on-chip memory, milliwatt power budgets, and real-time latency requirements, necessitating co-design approaches that tightly couple computation and memory access. FPGA-based accelerators are particularly effective in this regime due to their ability to implement custom dataflows that minimize intermediate storage. A representative example is the RISC-V-based TinyML accelerator employing fused pixel-wise dataflows for depthwise separable convolutions, which computes complete output pixels through tightly coupled pipelines without materializing intermediate feature maps \cite{yildirim2025risc}. This zero-buffer execution model substantially reduces data movement and achieves significant speedup and energy efficiency compared to layer-wise execution \cite{yildirim2025risc}.

Complementary approaches reduce training and adaptation costs. Green Granular Neural Networks replace iterative gradient descent in early network layers with FPGA-based direct equation solving, eliminating backpropagation overhead while enabling efficient incremental learning for higher layers \cite{Chu2023GGNN, vahdatpour2025energy}. To reduce engineering effort, automated co-design frameworks such as SECDA explore dataflow and memory configurations for FPGA accelerators, achieving notable energy savings, though current tools remain limited in their ability to support irregular models such as transformers \cite{Haris2021SECDA, kachris2025survey}.

Energy-harvesting TinyML systems face additional challenges due to highly variable power availability. ATM-Net addresses this by dynamically adapting precision and network depth based on instantaneous charging rate and stored energy, achieving large power reductions while maintaining competitive accuracy \cite{solanki2025atm}. SUSHI further manages latency through selective activation of subnetworks based on input complexity and quality-of-service constraints \cite{Behnam2023SUSHI}. These designs highlight a fundamental trade-off between specialization and flexibility: highly specialized accelerators maximize efficiency for narrow workloads, while more flexible frameworks sacrifice peak efficiency to support a broader range of models \cite{yildirim2025risc, Haris2021SECDA}.

\subsection{Mid-Scale: Edge-Cloud Split Computing}

Edge--cloud architectures distribute computation between local devices and remote servers to balance latency, energy consumption, and resource utilization \cite{may2024dynasplit}. DynaSplit formulates this problem as a multi-objective optimization over neural network partitioning and hardware configurations, combining offline exploration of Pareto-optimal solutions with online adaptation to quality-of-service requirements \cite{may2024dynasplit}. Experimental results demonstrate that split execution can substantially reduce total energy consumption compared to cloud-only inference, although benefits depend strongly on model architecture and workload characteristics \cite{may2024dynasplit}.

Smaller models optimized for mobile deployment often achieve highest efficiency when executed entirely at the edge, as communication overhead outweighs computational savings, whereas larger convolutional and transformer-based models benefit more from split execution \cite{may2024dynasplit}. Hardware-level parameters further complicate the design space: edge accelerators can reduce total energy despite higher instantaneous power draw due to reduced execution time, and CPU frequency scaling exhibits non-monotonic energy–latency trade-offs \cite{may2024dynasplit}. These interactions, together with the sensitivity of split-layer selection, necessitate empirical exploration rather than closed-form modeling. Practical deployments also face challenges including configuration switching overheads and network variability, which are often not fully captured in offline optimization frameworks \cite{may2024dynasplit}. Vision-based edge applications such as air-quality forecasting illustrate how co-design enables deployment of statistically informed CNN pipelines on resource-constrained devices while preserving accuracy \cite{vahdatpour2025forecasting}.

\subsection{Large-Scale: Transformers and LLMs}

Transformers and large language models impose extreme computational and memory demands, motivating co-design strategies that reduce data movement and improve utilization. FPGA-based spatial accelerators offer energy-efficient alternatives to GPUs by employing customized dataflows, persistent on-chip storage, and tiling strategies for attention projections, demonstrating competitive throughput under tight power budgets \cite{li2025design, chen2024understanding}. Surveyed FPGA and ASIC-based transformer accelerators consistently rely on model compression, dataflow specialization, and architectural innovations to achieve energy efficiencies substantially higher than GPU-based baselines \cite{kachris2025survey}.

Compute-in-memory architectures directly address the memory wall by colocating computation with storage \cite{wolters2024memory}. Recent SRAM-based designs integrate matrix-vector multiplication and nonlinear operations such as softmax within the same memory structures, enabling concurrent execution and significant gains in throughput and energy efficiency compared to GPUs \cite{kim2025hardware}. Digital CIM approaches further improve efficiency through weight-stationary execution and sparsity exploitation, achieving high TOPS/W for attention workloads \cite{yu2025digital, ma2025hnm}. Processing-in-memory implementations within DRAM banks enable large speedups for attention computation, though practical deployment is constrained by precision, refresh overhead, and thermal limitations \cite{Zhou2022TransPIM, wolters2024memory}.

System-level optimizations play a critical role in deployment efficiency. Benchmarking across multiple LLM families shows that energy consumption is highly sensitive to task type, sequence length, and runtime behavior, and that techniques such as DVFS, batching, speculative decoding, and KV-cache management can significantly reduce inference energy without modifying model semantics \cite{maliakel2025investigating, fernandez2025energy}. Quantization enables execution of LLMs on edge devices previously incapable of running such models, though aggressive precision reduction increases sensitivity to numerical instability \cite{husom2025sustainable}. Holistic evaluation incorporating datacenter efficiency metrics reveals that operational energy and infrastructure costs, including cooling and water usage, contribute substantially to the overall environmental footprint of large-scale deployments \cite{jegham2025hungry}.

Beyond conventional accelerators, specialized platforms explore alternative co-design points. Wafer-scale systems eliminate off-chip memory access for selected workloads through extreme spatial integration, achieving unprecedented training throughput at the cost of increased complexity and limited accessibility \cite{Lie2023Cerebras}. Mixture-of-experts architectures combined with FPGA acceleration activate only task-relevant subnetworks to reduce computation while preserving accuracy \cite{Fan2022M3ViT}. Datacenter-scale co-design efforts coordinate compilers, communication protocols, and hardware configurations to improve training efficiency across large clusters \cite{FireFlyer2024}, while SmartNIC-based accelerators extend co-design into the network fabric for memory-bound inference workloads \cite{Guo2023SmartNIC}.

\section{Cross-Scale Analysis and Challenges}
\label{sec:analysis}

\subsection{Cross-Scale Comparison}

Table~\ref{tab:comparison} summarizes representative energy-efficiency results across the TinyML-to-LLM spectrum. Across scales, FPGA-based implementations consistently achieve higher energy efficiency than GPU baselines, particularly for transformer workloads where customized dataflows reduce memory traffic \cite{kim2025hardware, li2025design, yildirim2025risc}. For example, HASTILY achieves 16--36$\times$ higher energy efficiency than an Nvidia A40 GPU by accelerating softmax within compute-in-memory (CIM) structures \cite{kim2025hardware}, while the RISC-V TinyML accelerator demonstrates a 59$\times$ speedup by eliminating intermediate memory accesses through zero-buffer execution \cite{yildirim2025risc}. However, GPUs retain an advantage in absolute throughput for large-batch inference, indicating that FPGA benefits are most pronounced in latency-sensitive and power-constrained deployments \cite{kachris2025survey}. ASIC designs offer the highest efficiency through specialization, but limited reconfigurability and high cost restrict their applicability; wafer-scale systems such as Cerebras WSE-2 exemplify this trade-off, achieving exceptional throughput at the expense of deployment flexibility \cite{Lie2023Cerebras}.

Energy efficiency does not scale monotonically with model size or deployment setting. In edge--cloud scenarios, smaller models such as MobileNetV2 achieve optimal efficiency through edge-only execution, whereas larger architectures (e.g., VGG16 and Vision Transformers) benefit from split computing \cite{may2024dynasplit}. For LLMs, quantization exhibits task-dependent behavior: aggressive 4-bit precision preserves accuracy for language understanding tasks but degrades performance on arithmetic reasoning \cite{husom2025sustainable}. System-level optimizations, including DVFS and inference batching, provide substantial energy reductions (30--73\%), yet interact non-linearly with workload characteristics such as batch size, sequence length, and decoding strategy \cite{maliakel2025investigating, fernandez2025energy}. Across all scales, memory access remains the dominant contributor to energy consumption—TinyML zero-buffer designs reduce data movement by up to 87\% \cite{yildirim2025risc}, CIM architectures eliminate off-chip transfers \cite{kim2025hardware, yu2025digital}, and split computing trades network communication against local memory access \cite{may2024dynasplit}. This consistency highlights memory-centric optimization as a unifying principle, despite substantial variation in architectural realization.

Quantitatively, these trends are consistent across deployment scales. TinyML 
FPGA-based approaches achieve 3.5--59$\times$ speedup with 87--88\% data 
movement reduction compared to software baselines \cite{yildirim2025risc, 
Haris2021SECDA}. For LLMs, CIM architectures demonstrate 16--36$\times$ 
energy efficiency gains over GPUs \cite{kim2025hardware, yu2025digital}, 
while system-level optimizations provide 30--73\% energy reduction through 
DVFS and inference serving techniques \cite{fernandez2025energy, 
maliakel2025investigating}.

\begin{table}[t]
\centering
\caption{Cross-Scale Energy Efficiency Comparison. Representative works demonstrate FPGA and CIM advantages for power-constrained deployments, with efficiency gains of 16-59× over baseline implementations.}
\label{tab:comparison}
\scriptsize
\begin{tabular}{|l|l|c|c|c|}
\hline
\textbf{Work} & \textbf{Platform} & \textbf{Model} & \textbf{Speedup} & \textbf{Key Metrics} \\
\hline
\multicolumn{5}{|c|}{\textit{TinyML Scale}} \\
\hline
\cite{yildirim2025risc} & FPGA (28nm) & DSC & 59.3$\times$ & 0.28mm$^2$ \\
\cite{solanki2025atm} & FPGA (32nm) & DenseNet & 87.5\% pwr$\downarrow$ & 0.14J PDP \\
\cite{Chu2023GGNN} & FPGA & Custom & Training$\uparrow$ & Energy$\downarrow$ \\
\hline
\multicolumn{5}{|c|}{\textit{Edge-Cloud Scale}} \\
\hline
\cite{may2024dynasplit} & RPi4+Cloud & VGG16/ViT & 72\% energy$\downarrow$ & 90\% QoS \\
\cite{vahdatpour2025forecasting} & Edge CNN & Custom & --- & Low power \\
\hline
\multicolumn{5}{|c|}{\textit{LLM Scale}} \\
\hline
\cite{kim2025hardware} & CIM (28nm) & BERT & 4.4--9.8$\times$ & 16--36$\times$ vs GPU \\
\cite{yu2025digital} & CIM (65nm) & Attention & 25$\times$ vs CPU & 34.1 TOPS/W \\
\cite{li2025design} & FPGA (KV260) & DistilBERT & 7$\times$ vs ARM & 3.1 GFLOPS \\
\cite{maliakel2025investigating} & GPU (DVFS) & LLaMA/Falcon & --- & 30\% energy$\downarrow$ \\
\cite{husom2025sustainable} & RPi4 (4GB) & Quantized LLM & --- & Edge viable \\
\cite{fernandez2025energy} & GPU (vLLM) & GPT/LLaMA & --- & 73\% energy$\downarrow$ \\
\hline
\end{tabular}
\end{table}

\subsection{Identified Gaps and Challenges}

A key limitation of current co-design approaches is their poor transferability across scales. Techniques optimized for TinyML, such as zero-buffer dataflows, do not generalize to transformer attention with irregular access patterns \cite{yildirim2025risc, kim2025hardware}, while LLM-specific optimizations such as KV-cache management and speculative decoding are infeasible on memory-constrained edge devices \cite{fernandez2025energy}. Design space exploration remains costly and fragmented: DynaSplit requires evaluating thousands of configurations to identify Pareto-optimal solutions \cite{may2024dynasplit}, and FPGA toolchains such as DNNBuilder are largely restricted to CNN architectures, limiting applicability to emerging transformer-based models \cite{Zhang2018DNNBuilder, kachris2025survey}. Although LLM-assisted co-design has emerged as a promising direction, it currently relies on expert supervision and lacks robust validation pipelines \cite{Yan2023LLMCoDesign}.

Benchmarking practices further limit practical insight. Many evaluations assume idealized conditions, such as stable network connectivity and static cloud resources, which do not reflect production environments \cite{may2024dynasplit}. Energy measurements frequently rely on software estimation rather than hardware-level profiling, leading to systematic underestimation of real-world consumption \cite{husom2025sustainable, fernandez2025energy}. Reconfiguration overheads, such as hardware switching latency on the order of hundreds of milliseconds, can violate quality-of-service constraints in time-critical applications \cite{may2024dynasplit}. Moreover, throughput-centric metrics obscure broader sustainability impacts; infrastructure-aware analyses show that environmental cost extends beyond energy to include water usage and embodied carbon, and that FLOP-efficient models may still incur high operational emissions due to inefficient memory behavior \cite{jegham2025hungry, fernandez2025energy}. These observations underscore the need for standardized lifecycle-aware benchmarking frameworks.

Architectural innovation introduces additional challenges. CIM efficiency depends strongly on memory technology: SRAM-based designs provide high precision but limited density \cite{kim2025hardware, yu2025digital}, while emerging memories promise higher density at the cost of endurance, drift, and write-energy constraints \cite{wolters2024memory, Zhou2022TransPIM}. Rapid algorithmic evolution further complicates hardware specialization; alternative transformer variants and state-space models alter computational patterns \cite{Guo2025LLMSurvey}, while Mixture-of-Experts architectures introduce dynamic sparsity poorly matched to static hardware \cite{Fan2022M3ViT}. Finally, aggressive efficiency optimizations introduce new security and robustness concerns, as reduced-precision arithmetic and custom dataflows expand the attack surface for side-channel leakage and adversarial exploitation \cite{Zhang2023RobustCoDesign}. Recent work on secure co-design highlights the necessity of integrating security considerations early in the design process \cite{Dubey2023SecureInference}, yet systematic frameworks for balancing efficiency, adaptability, and security remain largely unexplored \cite{Banitaba2025Stochastic}.

\section{A HIERARCHICAL PERSPECTIVE ON GREEN CO-DESIGN}
\label{sec:position}
In this section, we take a position-oriented perspective and synthesize insights from prior sections to reason about cross-scale energy-efficient co-design.

\subsection{Bridging Scales with Hierarchical Decomposition}

Conventional deep learning pipelines apply uniform optimization across all layers, despite their heterogeneous functional roles. Prior work has argued for hierarchical decomposition as a means of separating neural networks into two tiers: lower layers responsible for stable feature extraction and upper layers performing adaptive reasoning \cite{Chu2023GGNN, vahdatpour2025energy}. In this formulation, the lower tier employs FPGA-based direct equation solving, $W = (X^T X + \lambda I)^{-1} X^T Y$, eliminating backpropagation and enabling single-step convergence with reduced energy consumption \cite{Chu2023GGNN}. The upper tier applies incremental learning, $\theta_{H}^{(t+1)} = \theta_H^{(t)} + \eta \nabla_{\theta_H} \mathcal{L}$, with Elastic Weight Consolidation mitigating catastrophic forgetting, $\mathcal{L}_{EWC} = \mathcal{L} + \frac{\lambda}{2}\sum_i F_i(\theta_i - \theta_i^*)^2$, as explored in prior work \cite{vahdatpour2025energy}. Aligning optimization strategies with computational roles helps address cross-scale generalization limitations observed in existing co-design techniques.

\subsection{From TinyML to LLMs: A Unified Compound Framework}

This hierarchical perspective extends naturally to large language models through a Compound LLM architecture \cite{vahdatpour2025energy}. A compact, FPGA-optimized lower-tier LLM performs energy-efficient representation extraction suitable for edge deployment \cite{solanki2025atm, husom2025sustainable}, while an upper-tier LLM applies selective incremental updates to task-relevant subnetworks rather than retraining the full parameter space. This design enables energy efficiency via hardware acceleration of stable computations, flexibility through adaptive learning in upper layers, and scalable deployment across heterogeneous platforms \cite{may2024dynasplit}.

The framework integrates techniques surveyed in Section~\ref{sec:survey}: multi-precision arithmetic enables tier-specific quantization \cite{solanki2025atm}, zero-buffer dataflows minimize lower-tier memory traffic \cite{yildirim2025risc, kim2025hardware}, compute-in-memory accelerates attention in upper tiers \cite{yu2025digital, ma2025hnm}, and hybrid N:M sparsity applies structured pruning to stable layers while preserving task adaptability \cite{ma2025hnm}. Open research questions include automated tier boundary selection, theoretical convergence guarantees under hybrid optimization, and extension to trillion-parameter models, but hierarchical decomposition offers a principled path toward scalable and energy-efficient learning across deployment scales \cite{wolters2024memory}. Rather than targeting a specific instantiation, this perspective aims to clarify design trade-offs and identify principled directions for future software--hardware co-design.

\section{FUTURE WORK}
\label{sec:future}

The cross-scale analysis in this survey highlights several directions for advancing green AI systems capable of supporting complex big data applications. At the model level, emerging architectures that reduce memory complexity and enable conditional computation offer opportunities for improved energy efficiency, but realizing their benefits will require hardware and compiler support that can adapt to irregular execution patterns. Hybrid platforms that integrate reconfigurable logic, compute-in-memory, or neuromorphic components may provide complementary strengths for stable feature extraction, temporal reasoning, and continuous adaptation.

At the system level, future research should move beyond isolated device optimization toward holistic co-design that accounts for communication, storage, and orchestration costs in distributed and edge--cloud environments. Techniques such as near-data processing, hierarchical aggregation, and energy-aware task placement are particularly relevant for large-scale data pipelines where data movement dominates energy consumption. Automated co-design tools and benchmarking methodologies that reflect real-world constraints rather than idealized laboratory settings remain open challenges.

More broadly, progress toward sustainable AI will require evaluation frameworks that extend beyond raw performance and operational energy to incorporate lifecycle considerations, including resource utilization and environmental impact. Addressing these challenges will enable co-design methodologies that scale from resource-constrained edge devices to foundation-model infrastructures, supporting energy-efficient learning and inference for increasingly complex data-driven applications.

\section{CONCLUSION}
\label{sec:conclusion}

This work examined energy-efficient software--hardware co-design across the full machine learning spectrum, from milliwatt-scale TinyML systems to megawatt-class LLM infrastructures. Across deployment contexts, memory access—not arithmetic computation—emerges as the dominant energy bottleneck. FPGA-based accelerators reduce data movement through custom dataflows, compute-in-memory architectures further mitigate memory overhead by collocating storage and computation, and system-level techniques substantially lower LLM inference energy through batching and dynamic voltage--frequency scaling. Despite these advances, efficiency techniques remain fragmented: solutions optimized for CNN-based TinyML do not readily transfer to transformers, and datacenter-scale optimizations are often infeasible for edge deployment.

Persistent challenges include the high cost of design space exploration, limited automation in co-design toolchains, and evaluation methodologies that fail to reflect real-world operational and environmental constraints. Moreover, sustainability assessments frequently overlook lifecycle factors such as embodied carbon and water usage. Addressing these gaps requires reasoning about energy efficiency across heterogeneous computational roles rather than applying uniform optimization strategies throughout a model. A hierarchical view that aligns distinct optimization mechanisms with stable feature extraction and adaptive reasoning provides a natural lens for bridging deployment scales, from resource-constrained edge devices to large-scale foundation models. Achieving sustainable artificial intelligence will therefore require continued co-evolution of algorithms, hardware, and evaluation practices, with energy efficiency treated as a first-class design objective alongside accuracy and performance. As AI systems continue to scale and proliferate, co-design methodologies that bridge algorithmic innovation with architectural specialization will be essential for sustainable deployment.

\bibliographystyle{IEEEtran}
\bibliography{main}

\end{document}